\documentclass[10pt]{article}
\usepackage{graphicx,floatflt,amssymb,epsfig,epsf} 
\def\be {\begin{equation}}
\def\ee {\end{equation}}
\def\bea {\begin{eqnarray}}
\def\eea {\end{eqnarray}}

\def\g {\gamma}

\def\ztwo {{\cal Z}_2}

\def\stwo {\sqrt{2}}

\def\opcit(#1){ {\em op. cit.}, #1}
\def\etal {\em et al.}

\def\issue(#1,#2,#3){#1 (#3) #2} 

\def\APP(#1,#2,#3){Acta Phys.\ Polon.\ \issue(#1,#2,#3)}
\def\ARNPS(#1,#2,#3){Ann.\ Rev.\ Nucl.\ Part.\ Sci.\ \issue(#1,#2,#3)}
\def\CPC(#1,#2,#3){Comp.\ Phys.\ Comm.\ \issue(#1,#2,#3)}
\def\CIP(#1,#2,#3){Comput.\ Phys.\ \issue(#1,#2,#3)}
\def\EPJC(#1,#2,#3){Eur.\ Phys.\ J.\ C\ \issue(#1,#2,#3)}
\def\EPJD(#1,#2,#3){Eur.\ Phys.\ J. Direct\ C\ \issue(#1,#2,#3)}
\def\IEEETNS(#1,#2,#3){IEEE Trans.\ Nucl.\ Sci.\ \issue(#1,#2,#3)}
\def\IJMP(#1,#2,#3){Int.\ J.\ Mod.\ Phys. \issue(#1,#2,#3)}
\def\JHEP(#1,#2,#3){J.\ High Energy Physics \issue(#1,#2,#3)}
\def\JPG(#1,#2,#3){J.\ Phys.\ G \issue(#1,#2,#3)}
\def\MPL(#1,#2,#3){Mod.\ Phys.\ Lett.\ \issue(#1,#2,#3)}
\def\NP(#1,#2,#3){Nucl.\ Phys.\ \issue(#1,#2,#3)}
\def\NIM(#1,#2,#3){Nucl.\ Instrum.\ Meth.\ \issue(#1,#2,#3)}
\def\PL(#1,#2,#3){Phys.\ Lett.\ \issue(#1,#2,#3)}
\def\PRD(#1,#2,#3){Phys.\ Rev.\ D \issue(#1,#2,#3)}
\def\PRL(#1,#2,#3){Phys.\ Rev.\ Lett.\ \issue(#1,#2,#3)}
\def\SJNP(#1,#2,#3){Sov.\ J. Nucl.\ Phys.\ \issue(#1,#2,#3)}
\def\ZPC(#1,#2,#3){Zeit.\ Phys.\ C \issue(#1,#2,#3)}

\begin{document} 
\begin{flushright} 
CU-PHYSICS/15-2005\\
\end{flushright} 
\vskip 30pt 
 
\begin{center} 
{\Large \bf The International Linear Collider as a Kaluza-Klein Factory}\\
\vspace*{1cm} 
\renewcommand{\thefootnote}{\fnsymbol{footnote}} 
{\large {\sf Biplob Bhattacherjee} and {\sf Anirban Kundu} } \\ 
\vspace{10pt} 
{\small 
   {\em Department of Physics, University of Calcutta, 92 A.P.C. 
        Road, Kolkata 700009, India}}
 
\normalsize 
\end{center} 
 
\begin{abstract} 
In the minimal Universal Extra Dimension model, single production of $n=2$ 
gauge bosons provides a unique discriminating feature from supersymmetry.
We discuss how the proposed International Linear Collider can act as
a $n=2$ factory, much in the same vein as LEP. We also touch upon the potential
of the $\gamma\gamma$ mode of the collider to study the production and
the decay of an intermediate mass Higgs boson and its KK excitations.
 
\vskip 5pt \noindent 
\texttt{Keywords:~~Universal Extra Dimension, Electron-Positron Collider}
\end{abstract}

\renewcommand{\thesection}{\Roman{section}} 
\setcounter{footnote}{0} 
\renewcommand{\thefootnote}{\arabic{footnote}} 

\section{Introduction}

The possibility of a compactified extra dimension was first discussed
by Kaluza and Klein \cite{kaluza}, and such extra dimensional models
were later revived by the necessity of a consistent formulation of
string theories. There are a number of such models, and
they differ mainly in two ways: first, the number of extra dimensions, the
geometry of space-time, and
the compactification manifold, and second, which particles can go into
the extra dimensions (hereafter called bulk) and which cannot. 

We will focus on the so-called Universal Extra Dimension (UED) model
proposed by Appelquist, Cheng, and Dobrescu \cite{acd}. 
In this model, all SM particles
can go into the bulk. In the simplest UED scenario, there is only one
extra dimension, denoted by $y$, compactified on a circle ($S_1$) of
radius $R$. The model predictions
remain essentially unchanged if there are more than one extra dimension, 
with a hierarchical radii of compactification. Our discussion will be on
the simplest scenario only. 

To get chiral fermions at low-energy, one must impose a further $\ztwo$
symmetry ($y\leftrightarrow -y$), so that finally we have an $S_1/\ztwo$
orbifold. As is well-known, a higher dimension theory is nonrenormalisable
and should be treated in the spirit of an effective theory  valid upto
a scale $\Lambda > R^{-1}$. All fields have five space-time components;
when brought down to four dimensions, for each low-mass (zero-mode) Standard
Model (SM) particle of mass $m_0$, 
we get an associated Kaluza-Klein (KK) tower, the
$n$-th level (this $n$ is the KK number of the particle)
of which has a mass given by
\be
m_n^2 = m_0^2 + \frac{n^2}{R^2}.
  \label{kktree}
\ee
This is a tree-level relationship and gets modified once we take into
account the radiative corrections.

Another important feature of the UED scenario is the conservation of the
KK number. This is simply a reflection of the fact that all particles can go
into the fifth dimension and so the momentum along the fifth dimension
must be conserved. Also, this means that the lowest-mass $n=1$ particle,
which turns out to be the $n=1$ photon, is absolutely stable. Such a
lightest KK particle (LKP), just like the lightest supersymmetric particle
(LSP), is an excellent candidate for dark matter \cite{servant,mazumder,byrne,
kk2dark}. 

Radiative corrections to the masses of the KK particles have been computed
in \cite{georgi,cheng1,pk}. These papers, in particular \cite{cheng1}, show
that the almost mass-degenerate spectrum for any KK level, resulting from
eq.\ (\ref{kktree}), splits up due to such correction terms. There are two
types of correction; the first one, which results just from the compactification
of the extra dimension, is in general small (zero for fermions) and is
constant for all $n$ levels. This we will call the bulk correction. The second
one, which we will call boundary correction, is comparatively large (goes
as $\ln\Lambda^2$ and hence, in principle, can be divergent), and plays
the major role in  determining the exact spectrum and possible decay modes.
The boundary correction terms are related with the interactions present only
at the fixed points $y=0$ and $y=\pi R$. If the interaction is symmetric 
under the exchange of these fixed points (this is another $\ztwo$ symmetry,
but not the $\ztwo$ of $y\leftrightarrow -y$), the conservation of KK
number breaks down to the conservation of KK parity, defined as $(-1)^n$.
Thus, LKP is still stable, but it is possible to produce an $n=2$ state
from two $n=0$ states. This particular feature will be of central interest
to this letter.
 
The low-energy phenomenology has been discussed in \cite{acd,
agashe,chk,buras,papavas,oliver,ewued}, and the high-energy collider signatures
in \cite{rizzo,nandi1,nandi2,bdkr,asesh,cheng2,riemann}. 
The limit on $1/R$ from
precision data is about 250-300 GeV, while the limit estimated from dark
matter search \cite{servant} is about a factor of two higher. 
The loop corrections are quite insensitive
to the precise values of the radiative corrections. With the proposed reach
of ILC in mind, we will be interested in the range 300 GeV $<R^{-1}<$ 500
GeV. (The role of linear colliders in precision study of TeV-scale extra
dimension models has been emphasized in, {\em e.g.}, \cite{antoniadis}.) 

One of the reasons that UED has been considered seriously for the next 
generation collider experiments is the fact that in a certain region of the
parameter space, supersymmetric models can mimic the UED signals. In this
parameter space, sleptons are almost degenerate with electroweak gauginos
and the lightest neutralino is only slightly separated from the lighter
chargino (analogous to the anomaly-mediated supersymmetric models). If one
only observes the $n=1$ states, such a discrimination is difficult in LHC,
though how the spin of the excited state can be determined has been discussed
in the literature \cite{barr,webber}. At ILC or CLIC, the discrimination is
simpler; even if one looks for the production of $n=1$ electron \cite{bdkr}
or muon \cite{asesh} pair, the angular distribution of the ultimate soft
leptons will be a good discriminator; this is just based on the simple fact
that the decay distribution from a spin-0 object is different from a spin-1/2
one, whether or not there is a $t$-channel contribution.  

Let us mention here that though
the main focus is on the ILC, an identical study may be performed for
CLIC, the proposed multi-TeV $e^+e^-$ machine, with an optimised $\sqrt{s}
=3$ TeV (and may be upgraded to 5 TeV), and luminosity of $10^{35}$ cm$^{-2}$
s$^{-1}$. The electron beam at CLIC may be polarised upto 80\%,
and the positron beam upto 60-80\%, from Compton scattering off a high power
laser beam \cite{clic}. Clearly, the reach of CLIC will be much higher.

It has been pointed out \cite{cheng2} that a `smoking gun' signal of
UED would be the production of $n=2$ states. Pair production of such states
is difficult even at the LHC energy, and is surely out of reach for ILC.
However, one can produce a single $\g_2$ or $Z_2$ \cite{asesh,riemann}. 
These will be narrow
peaks, closely spaced, and probably not resolvable at LHC. (In fact, as we
will show later, due to the decay pattern of $\g_2$, almost entirely to
two jets plus no missing energy, it will be very difficult to locate this
resonance at LHC.) Here ILC will
perform a much better job, and if it can sit on these resonances, it may
even repeat the LEP-I story. Such precision measurements will definitely 
determine the model parameters, even if it is not the simplest UED model.  
There are a couple of points that the reader should note.
\begin{itemize}
\item
If a collider is energetic enough to pair produce $n=1$ excitations, single
production of $n=2$ states is also possible. Since it is not possible to
produce only one $n=1$ UED state, it is a none-or-both situation.  
\item 
Decay of a $n=2$ state to two $n=0$ states is allowed by KK parity conservation,
but this is suppressed by boundary-to-bulk ratio. However, there is no
phase space suppression, not even if the final state is a $t\bar{t}$ pair.
On the other hand, the coupling is large for the KK number conserving decays
($2\to 2-0,1-1$, where the numbers are for the generic KK levels), 
but there is a heavy kinematic suppression. Ultimately
it turns out that both suppressions are of equal importance \cite{cheng1} and
hence both KK conserving and KK violating decays are to be taken into
account.
\end{itemize} 

In this letter we will discuss the role that ILC may play in studying this
resonance physics. We will also mention how the production and the decay
of the Standard Model (SM) Higgs boson as well as its excitations can be
studied in the $\gamma\gamma$ mode of the collider. In particular we will
focus on the intermediate mass Higgs boson ($115$ GeV $< m_h<160$ GeV) case.
The decay that is most seriously affected is
$h\to gg$, the decay to a gluon pair. (Obviously, the SM Higgs production
in the gluon-gluon fusion channel at LHC will also show a marked change;
this has been discussed in \cite{petriello}.) The reason for such a change
is the higher $n$ particles running in the loop, but unlike the SM, the
effect is decoupling in nature, at least if there is only one extra dimension;
otherwise the result is divergent and one is forced to use some hard
truncation. The $h\to\g\g$ and $h\to\g Z$ vertices are comparably less affected.  

\section{The KK number violating interactions}

A consistent formulation of UED needs the inclusion of interaction terms that
exist only at the fixed points \cite{georgi,cheng1}. 
These terms can in principle be non-universal (just like non-universal soft
mass terms in supersymmetry) and hence can affect the FCNC processes, but
in the simplest UED model, they are taken to be universal, symmetric about
the fixed points, and vanishing for energy $\Lambda \gg R^{-1}$. This
introduces only two new parameters in the model, $\Lambda$ and $R^{-1}$,
and ensures the conservation of KK parity. (In fact, there is a third
parameter, ${\bar{m}_h}^2$, the Higgs mass term induced on the fixed
points. In the minimal UED model this is assumed to be zero, but its
precise value may be probed through a precision study as shown later.) 

The excited fermions are vectorial. For SU(2) doublets, the left-handed
components are $\ztwo$-even and right-handed components are $\ztwo$-odd,
and the opposite is true for SU(2) singlets. The Yukawa terms in the fermion
mass matrices may mix the doublet and the singlet states; however, this
is numerically significant only for the top quark. The first four components of
the gauge bosons are $\ztwo$-even, while the fifth one is a $\ztwo$-odd
scalar. A combination of this and the excitation of the zero-mode charged
Goldstone boson is absorbed by the excited gauge boson, and the orthogonal
combination remains in the spectrum as a physical charged Higgs. The Higgs
spectrum is nearly degenerate, with $m_{h_1} > m_{A_1} > m_{h_1^\pm}$.
To a very
good approximation, the charged Higgs is the excitation of the zero-mode
Goldstone. A complete set of Feynman rules is given by \cite{buras}.

The excited states of $Z$ and photon are obtained by diagonalising the 
mass matrix of $W_3$ and $B$. It has been shown in \cite{cheng1} that
for all practical purpose, the $n=2$ excitation of $Z$ is almost $W_3$
(so that it is a pure SU(2) triplet and couples only to the left-handed
fermions) while the $n=2$ excitation for photon is almost a pure $B$
(so that it couples with different strengths to left- and right-handed 
fermions). 

We will be interested in the coupling of $n=2$ gauge bosons with an
$n=0$ fermion-antifermion pair. This coupling is given by \cite{cheng1}
\be
\left(-ig\g^\mu T_a P_+\right) \frac{\stwo}{2} \left(
\frac{\bar\delta(m_{V_2}^2)}{m_2^2}-2\frac{\bar\delta(m_{f_2})}{m_2}\right),
   \label{defx}
\ee
where $g$ is the generic gauge coupling, $T_a$ is the group generator
(third component of isospin, or hypercharge), and $P_+$ is the 
$\ztwo$-even projection operator, which is $P_L=(1-\g_5)/2$ for $Z_2$,
but can be both $P_L$ or $P_R$ for $\g_2$. $V$ can be either $Z$ or $\g$.
The expressions for the
boundary corrections, $\bar\delta$, can be found in \cite{cheng1}.  
They are proportional to the renormalisation-group (RG) $\beta$-functions
times $\ln(\Lambda^2/\mu^2)$, where the regularisation scale $\mu$ may
be taken to be $2/R$ in this case. Clearly, in the fine-tuned case $\Lambda R
=2$, all KK number violating couplings are identically zero.

\begin{figure}
\vspace{-10pt}
\centerline{\hspace{-3.3mm}
\rotatebox{-90}{\epsfxsize=6cm\epsfbox{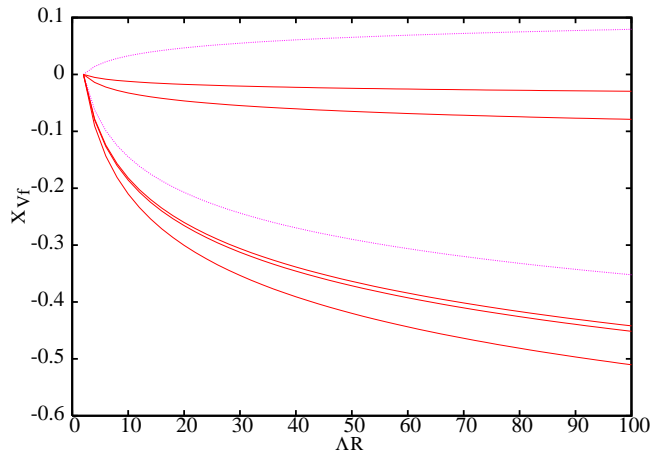}}}
\caption{$X_{Vf}$, the KK number violating couplings, as
a function of $\Lambda R$, for $R^{-1}=300$ GeV (the values are independent
of $R$). From top to bottom, the curves are for $X_{ZL}$, $X_{\g e}$,
$X_{\g L}$, $X_{ZQ}$, $X_{\g d}$, $X_{\g u}$, and $X_{\g Q}$ respectively.
For their definitions, see text.}
\end{figure}

It is easy to check that for any level, the excitation of the photon,
$\g_n$, is the lowest-lying particle. Thus, $\g_2$ cannot decay into a pair
of $n=0$ and $n=2$ fermions. In fact, the decay to an $n=1$ pair is also
kinematically forbidden, for all choices of $\Lambda$ and $R$. Thus, the only
possible way to decay is to an $n=0$ fermion-antifermion pair. Here, both
right- and left-handed pairs (of quarks and leptons, including neutrinos)
are included, albeit with different strengths, as obtained from eq.\ 
(\ref{defx}). In figure 1, we show how the function $X_{Vf}$, defined as
\be
X_{Vf}= \frac{\stwo}{2} \left(
\frac{\bar\delta(m_{V_2}^2)}{m_2^2}-2\frac{\bar\delta(m_{f_2})}{m_2}\right),
   \label{xvfdef}
\ee
varies for $V=\g,Z$ and $f=u_i,d_i,e_i$ (SU(2) singlet states) and $L_i,
Q_i$ (SU(2) doublet states), where $i$ is the generation index. It is
obvious that $\g_2$ should decay almost entirely to a $q\bar{q}$ pair, 
because of the larger splitting between $\g_2$ and $n=2$ quarks. Altogether,
there are 45 channels, including the colour degrees of freedom. 

One may ask whether $\g_2$ can decay into KK-number conserving three- or
four-body channels, {\em e.g}, $\g_2\to e_1^+ {e_1}^{-\ast} \to
e_1^+ e_0^- \g_1$. The answer is {\em no}, in particular for the minimal UED
model. The reason is that KK-number conserving decays must result in two
LKPs in the final state, and $2m_{\g_1} > m_{\g_2}$ over the entire parameter
space. This we have checked both analytically and numerically; in fact, an
analytical check is easy if one looks at the mass corrections and $W_3-B$
mixing effects \cite{cheng1}. 

The decay pattern of $Z_2$ is more complicated. It is an almost pure
$(W_3)_2$, so it couples only to left-handed doublet fermions. Kinematically,
decay to an $n=1$ pair of lepton doublet ($\ztwo$-even)
is allowed, except for very low
values of $\Lambda$ ($\Lambda R < 3$). There are 6 such channels, including
neutrinos. These states will ultimately decay
to the corresponding $n=0$ leptons, plus $\g_1$, the LKP, (even the $n=1$ 
neutrino can decay in this channel), so that the signature
will be a pair of soft leptons (for charged lepton channels) 
plus a huge missing energy (excited neutrinos, of course, will go undetected).
Fortunately, these final soft leptons should be detectable \cite{bdkr,asesh}. 
Similarly, $Z_2$ can decay to a pair of $n=2$ and $n=0$ doublet leptons.
Again, there are 6 channels, plus 6 CP-conjugate ones.
Both these modes are KK-number conserving, but there is an important
difference: while the coupling is the usual $g$ for the latter channels,
it is $g/\stwo$ for the former ones. This can be checked by integrating
the trigonometric terms dependent on the fifth coordinate $y$. 

Just like $\g_2$, $Z_2$ has its own share of KK-number violating modes,
but it can only decay to a left-handed pair. Since the lower limit on $R^{-1}$
is about 300 GeV, both these gauge bosons can decay even to the $n=0$
$t\bar{t}$ pair. However, KK-number conserving $Z_2$ decays to electroweak
bosons are forbidden from kinematic considerations.

Since KK parity is conserved, the $s$-channel Feynman diagrams have only
the gauge bosons ($\ztwo$-even) as mediators, not the $\ztwo$-odd scalar
which is the fifth component of the gauge boson. This statement is, of course,
independent of the gauge choice. Similarly, the final state particles must
be both KK even or both KK odd. However, we have just shown that only 
$\ztwo$-even states will be produced in the final state.

In the minimal UED model, ${\bar{m}_h}^2=0$, $Z_2$ cannot decay through
the Bjorken channel to $Z_1 h_1$, purely from kinematic considerations.
(The three-body channels, with a virtual $Z_1$ or $h_1$, will be even
more suppressed.)
However, if ${\bar{m}_h}^2 < 0$, all the Higgs masses will be lowered,
and one can just be able to produce a neutral CP-even Higgs excitation
through this channel. The decay channel of $h_1$ is dominantly a right-handed
$\tau$ pair (assuming the mixing in the $n=1$ level to be small) plus
LKP, and if the $\tau$s are soft enough, they may escape detection,
leading to an invisible decay mode of $h_1$. Of course, the vertex
$Z_2 W_1^\pm h_1^\mp$ does not exist.

\section{Production and decay of n = 2 neutral gauge bosons}

The gauge bosons are produced as $s$-channel resonances in $e^+e^-$
collision through KK-number violating couplings. This suppression brings
down the peak cross-section to an otherwise expected nanobarn level to about
35-45 pb for $Z_2$ and about 63 pb for $\g_2$ (for $R^{-1}=300$ GeV,
and the variation is due to that of $\Lambda$). For $R^{-1}=450$ GeV,
these numbers drop to 16-21 pb and 28 pb, respectively. The reason for
a higher production cross-section for $\g_2$ is its narrower width compared
to $Z_2$. However, it will be almost impossible to detect $\g_2$ at LHC 
since it decays almost entirely to two jets which will be swamped by the
QCD background, and moreover the resonance is quite narrow. $Z_2$ has a
better chance, since there are a number of hadronically quiet channels, and
soft leptons with energy greater than 2 GeV should be detectable. But for
a precision study of these resonances we must turn to ILC (or CLIC). These
machines should be able to measure precisely the positions and the widths
of these two peaks, and hence entirely determine the spectrum, since there
are only two unknown parameters (hopefully the Higgs mass will already
be measured by LHC). These measurements, in conjunction with the precise
determination of $n=1$ levels, should be able to discriminate, not only
between UED and supersymmetry, but even the minimal version of UED from its
variants.

\begin{figure}
\vspace{-10pt}
\centerline{\hspace{-3.3mm}
\rotatebox{-90}{\epsfxsize=6cm\epsfbox{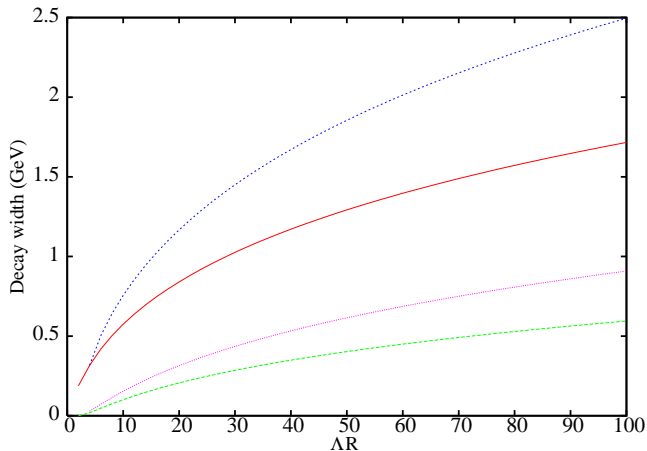}}}
\caption{Decay widths of $Z_2$ (upper pair) and $\g_2$ (lower pair)
as a function of $\Lambda R$, for $R^{-1}=450$ GeV and 300 GeV (upper and 
lower curves in a pair).}
\end{figure}

In figure 2 we show the decay widths of $Z_2$ and $\g_2$, plotted for
two different values of $R^{-1}$ and as a function of $\Lambda R$. They
increase logarithmically, because of the $\log\Lambda^2$ dependence of
the couplings, but no new channel opens up. For small values of $\Lambda R$
(2-3), the KK-number conserving channels for $Z_2$ are still closed, and 
$Z_2$ can be very long-lived, even to leave a displaced vertex. (As discussed
earlier, for $\Lambda R=2$, a somewhat fine-tuned value, $Z_2$ is almost
stable, and the peak is correspondingly narrow and hence difficult to
detect.)

We emphasize that this study will be meaningful only if LHC finds some
signal of new physics, which may look like UED, and for which the pair
production of $n=1$ states is not beyond the reach of ILC. In that case
a careful scan about $\sqrt{s}=2/R$ should reveal these two peaks. The points
that one would like to verify are:
\begin{itemize}
\item On the $Z_2$ peak, ${\cal R}$, the ratio of $e^+e^-$ to two jets to
$e^+e^-\to\mu^+\mu^-$ would show a sharp dip, in particular if we include
the missing energy events. The reason is that the $Z_2$-width is dominated by
the channel to a pair of $n=1$ leptons, and quarks can appear only from 
KK-number violating interactions. On the other hand, ${\cal R}$ should show
a sharp peak on the $\g_2$ resonance.
\item The cross-section would show a kink between the two peaks; this is
the position where the KK-number conserving channels open up.
\item With the polarised beam option, the behaviour of the two peaks will
be quite different. Since $Z_2$ couples only to the left-handed fermions,
with suitable polarisation the peak may vanish altogether, or may get
enhanced by a factor of 3 (assuming 80\% $e^-$ polarisation and 60-70\%
$e^+$ polarisation). The $\g_2$ peak will get enhanced by about a factor
of 2 with left-polarised $e^-$ beam, but will never vanish altogether.
The reason is that the hypercharge gauge boson $B$ couples to both $R$ and
$L$ fermions, and though $Y_{e_R}=2Y_{e_L}$, the ratio $X_{\gamma L}/
X_{\gamma e}$ (see eq.\ (\ref{xvfdef})) overcompensates it. 
\item As we have pointed out earlier, $\g_2$ cannot have a KK-number 
conserving decay channel. This is true only in the minimal UED model.
In a nonminimal version, there may be two possible corrections. First, due
to nonuniversal boundary terms, $\g_2$ may become a little more massive,
which will open the KK-number conserving windows. There is a chance that
this will spoil the nice feature of having the LKP as a viable dark matter
candidate. The second point, which is more probable, is to have asymmetric
boundary terms (different for $y=0$ and $y=\pi R$). This will break the 
KK-parity, and will result in decay modes like $\g_2\to e_1^+e_0^-$. Even
if such couplings are suppressed, a precision study at ILC may discriminate
between different models of UED.
\end{itemize}

Let us also note that the SM background, coming from the continuum, is
less than 10 pb for $\sqrt{s}=600$-900 GeV \cite{jlc}, and may be
further reduced by suitable cuts. 

\section{Production and decay of the Higgs boson and its excitations}

We focus on the production channel $\g\g\to h$, assuming that ILC will
have a $\g\g$ option. The importance of this channel has been emphasized
in the literature, particularly for intermediate mass Higgs.
The reason is that the production is loop-mediated,
and the excited states can run in the loop, without the need of a KK-number
violating vertex. 
(This is analogous
to the case of R-parity conserving supersymmetry.) Again, we assume the
Higgs to be in the intermediate mass range, 115 GeV$<m_h<$ 160 GeV.  

In the SM, the process $\g\g\to h$ proceeds mainly through the top quark
and the $W$-boson loops. The leading order expressions are to be found in
\cite{voloshin,spira}. The UED case, for a larger range of $m_h$ and $R^{-1}$
than we discuss, has been treated in detail in \cite{petriello}, and we
agree with the findings. Let us try to see the result in a physically more
transparent way.
\begin{itemize}
\item There are higher $n$ excitations in the loop. Only three loops are of
relevance: those of the top quark, the $W$ boson and the charged Higgs.
Though the masses of other fermionic excitations are large too, their
contributions get suppressed from their couplings with Higgs, which is
proportional to their zero-mode masses $m_f$. Moreover, in the fermionic loops,
there is one more chirality flip, which brings in another factor of $m_f$.
Note that the vectorial mass term cannot come since $\ztwo$-parity of the
virtual fermion cannot change inside the loop. Also, unlike SM, the $W$-loop
and the top quark loop interfere constructively.
\item For the $W$ loop and the charged Higgs loop, the mass suppression in the
vertex exists, but there is no question of a chirality flip. On the other
hand, if one wishes to normalise the result with the same prefactor of $G_F$
as in the SM, this brings another factor of $m_W/m_{W_1}$, thus making the
effective suppression quadratic in $vR$, where $v$ is the vacuum expectation
value of the SM Higgs field. (Note that $G_F\propto v^{-2}$).
In the unitary gauge, the fifth
component of $W^+$ does not appear. The top quark loop is also quadratically
suppressed in $vR$, but there are two such loops, due to the vectorial
nature of the excitation. 
\item With these modifications, one may use the standard expressions (leading
order is sufficient) for these loops, with a suitable scaling of the mass
of the loop particle. Note that all the amplitudes must be coherently added.
\end{itemize}

\begin{figure}
\vspace{-10pt}
\centerline{\hspace{-3.3mm}
\rotatebox{-90}{\epsfxsize=6cm\epsfbox{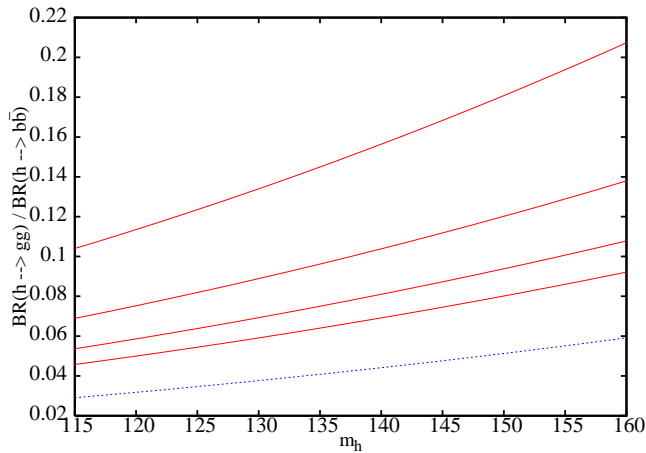}}}
\caption{The relative enhancement of $h\to gg$ decay width for
$R^{-1}=300$, 400, 500, and 600 GeV (top to bottom). The SM expectation is
the blue line at the bottom.}
\end{figure}

The result is in conformity with \cite{petriello}: the $\g\g\to h$ cross-section
drops by 25-30\% for $R^{-1}=300$ GeV, and by a smaller amount for higher
values of $R^{-1}$. The SM $W$-loop contribution is large and negative, but
the UED bosonic loops are positive. However, because of the quadratic 
suppression, they can only partly offset the SM contribution. 
The intermediate mass Higgs decays dominantly to $b\bar b$, or to three-body
final states through virtual gauge bosons \cite{carena}. Such tree-level
modes will hardly feel the effect of UED. But the loop-induced decay
$h\to gg$, a pair of gluons (second in importance in the intermediate mass
range), will be enhanced by UED (only the top quark loop is
relevant here). In figure 3 we show how the relative importance of this
mode changes with the dominant mode $h\to b\bar{b}$. The result is
convergent for one extra dimension.
We note that in the clean atmosphere of ILC, such
an enhancement will be easily detectable. A similar calculation, performed
in the context of LHC for the production of Higgs through gluon-gluon
fusion, shows an identical effect \cite{petriello}. 

ILC is not the ideal machine to produce the excited neutral Higgses. The best
channel is the fusion of a $W_0$ and a $W_1$, accompanied by a $\nu_0$ and
a $\nu_1$. The signal will be a soft $\tau$ pair, with a huge amount of
missing energy ($\nu_1$ decays to $\nu_0$ and LKP, and hence acts as a 
virtual LKP). The charged Higgs may be pair produced in the $\g\g$ mode
with a large cross-section. In the minimal UED model, ${\bar m}_h^2=0$,
$h_1^+$ can decay to a right-handed $\tau_1$ and a zero-mode $\nu_\tau$, and
the former will again give a soft $\tau$. If it is too soft, even the
charged Higgs may decay invisibly. Anyway, it is important to have a high
soft-$\tau$ detection efficiency (preferably $<1$ GeV) in the ILC detector.
It will be even better if one can measure the polarisation of the $\tau$.
The excited Higgs sector will be discussed in detail in a subsequent
publication.

\section{Summary}

ILC can act as a Kaluza-Klein factory if $R^{-1}$ happen to be on the lower
side so that one is able to pair produce $n=1$ states of the minimal UED
(otherwise we have to turn to CLIC). ILC can make a precision study of
the two peaks, $Z_2$ and $\g_2$, including precise measurements of their
positions and widths. The peaks are well over the continuum background,
and if the machine can be tuned properly, it can repeat the LEP-I story.
Precision study of these peaks will not only discriminate this model from
any supersymmetric scenarios that may mimic UED at the LHC, but also
be able to determine the model parameters $R$ and $\Lambda$, and hence
can potentially test the minimality of the UED version. The only parameter
that the resonance study may not be able to fix is ${\bar m}_h^2$, which
in turn may be determined through the study of the Higgs sector. The
$h\to gg$ mode should show a marked enhancement. On the other hand, the excited
Higgses will mostly decay to soft $\tau$s, and it is of utmost importance
to have a high soft-$\tau$ detection efficiency.

\centerline{\bf Acknowledgements}

We thank Anindya Datta, Santosh Kumar Rai, and Amitava Raychaudhuri,
and also the participants of the Study Group
on Extra Dimensions at LHC, held at HRI, Allahabad, for a number of
useful and stimulating discussions. A.K.\ thanks the Department of
Science and Technology, Govt.\ of India, for the research project
SR/S2/HEP-15/2003. B.B.\ thanks UGC, Govt.\ of India, for a research
fellowship.

\end{document}